# In-situ dynamic spatial reconfiguration of nanoplasmonics using photothermal-shock tweezers

*Runlin Zhu, Zhaoqi Gu, Tianci Shen, Yifei Liu, Zhangxing Shi, Shuangyi Linghu and Fuxing Gu\**

R. Zhu, Z. Gu, T. Shen, Y. Liu, Z. Shi, Prof. F. Gu
Laboratory of Integrated Opto-Mechanics and Electronics
School of Optical-Electrical and Computer Engineering
University of Shanghai for Science and Technology
Shanghai 200093, P. R. China
E-mail: gufuxing@usst.edu.cn

S. Linghu
Department of Physics
Guizhou University
Guiyang 550025, China

**Abstract**

Dynamic reconfiguration is crucial for nanoplasmonic structures to achieve diversified functions and optimize performances; however, the dynamic reconfiguration of spatial arrangements remains a formidable technological challenge. Here, we showcase in-situ dynamic spatial reconfiguration of plasmonic nanowire devices and circuits on dry solid substrates, by harnessing a photothermal-shock tweezers platform. Owing to its versatility, nanoscale precision, real-time operation, and large external output force, the multimodal platform enables dexterous fine-tuning of positions, overlap lengths, and coupling distances and orientations of discrete components in situ. Spatial position-dependent optical properties that have not been reported before or are challenging to achieve through traditional micro/nanomanipulation are easily tuned and observed, such as the intensity evolution of axial photon-plasmon coupling from near field to far field, and the resonant mode evolution of photonic cavity-plasmonic cavity coupling from weak to strong. We also employ the nanorobotic probe-based operation mode to optimize the side-mode suppression ratios of single-mode lasers and the intensity splitting ratios of 3-dB couplers. Our results are general and applicable to materials of almost any size, structure, and material type, as well as other narrow or curved micro/nano-waveguide surfaces, which opens new avenues for reconfigurable nanoplasmonic structures with dynamically tunable spatial features.

**1. Introduction**

Due to their high crystallinity and smooth surfaces, free-standing metal nanowires (NWs) synthesized via bottom-up methods (e.g., chemical seeding and thermal evaporation) exhibit relatively low propagation losses of surface plasmon polaritons (SPPs) while maintaining ultra-tight filed confinement, which dramatically enhances light-matter interaction and provides prominent applications in lasing, sensing, high-density integrated circuits, and surface-enhanced Raman scattering, luminescence and nonlinear optics.[1–6] Although marked advancements have been made in the fabrication of integrated nanoplasmonic devices and circuits, including couplers, emitters, beam splitters, and Mach-Zehnder interferometers, they are limited in their functions due to the static factory design and fabrication. That is, their optical properties are inherently fixed once constructed, and are difficult to adjust dynamically, which is crucial for achieving dynamic diversified functions such as tunability, switching and modulation of electromagnetic waves, and optimized performances. Over the past few decades, many approaches have been demonstrated to dynamically reconfigure the optical properties of plasmonic nanostructures through various means, such as electricity, heat, light, magnetism, and force.[7–9] However, dynamically reconfiguring the spatial structural arrangements between plasmonic discrete components and other functional groups is a formidable technological challenge, mainly due to the strong adhesion force between micro/nano-objects and solid substrates (up to ~μN level in dry air).[10–12] Despite advances in mechanical deformation[13-18] and DNA assembly,[19] the applicable conditions are mainly localized surface plasmon resonances and are limited by material types and spatial tuning distances, and it is also difficult to control individual target objects. For SPP NW-based nanostructures, dynamic reconfiguration of their spatial arrangements, such as tuning the positions, overlap lengths, coupling distances, and orientations, remains elusive and has not been reported.

The key to addressing this challenge is to develop micro/nano-manipulation technologies that can output strong thrust to overcome the strong adhesion force and also have high precision, versatility, dexterity, fast response, and intelligence. Currently, most SPP NW-based devices and circuits rely on manual assembly using contact probes to manipulate the discrete components,[10,20–22] including atomic force microscopy (AFM), and various micro/nanotips (e.g., silica fiber probes and tungsten probes). However, these contact methods present several drawbacks: 1. Sample surface damage may result from direct contact. 2. Axial and rotational manipulation along the NW axis is difficult. 3. Contact methods are challenging in unique scenarios involving liquids, solid-liquid interfaces, and low temperature or vacuum. 4. The inability to image and manipulate simultaneously hinders the real-time and fast operation of AFM technologies. In parallel, non-contact optical methods such as optical tweezers,[12,23–25] plasmonic tweezers,[26] and optoelectronic tweezers[27] have been proposed; however, they generate small forces (~pN level), so they predominantly operate in liquid suspension environments, but liquid convection, thermal perturbation, and surface tension can adversely affect manipulation accuracy and stability.

More recently, by using pulsed lasers, photothermal-shock tweezers (PTSTs) have been demonstrated to trap and manipulate micro/nano-objects on solid substrates,[28–31] with advantages of versatility, dexterity, high nanoscale precision, real-time operation, and large external output force. Here, we demonstrate in-situ dynamic reconfiguration of spatial arrangements of plasmonic NW-based devices and circuits on dry solid substrates by harnessing the multiple PTSTs platform. The direct operation modes, including axial and lateral transitions and lateral rotation, and the indirect operation mode of nanorobotic probes, enable dexterously manipulating single metal NWs and dynamically fine-tuning the coupling length ($L$), distance ($d$), and angle ($\theta$) between NWs and light-emitting sources in situ. Notably, we realize the intensity evolution of the photon-plasmon coupling along the axial direction from near field to far field, and the resonant mode evolution of the photonic cavity-plasmonic cavity coupling from weak to strong. Finally, we employ the nanorobotic probe-based operation mode to optimize the side-mode suppression ratios of single-mode hybrid photon-plasmon lasers and the intensity splitting ratios of 3-dB plasmonic couplers from two touching metal NWs.

## 2. Results and Discussion

Our PTST is a multimodal manipulation platform (Methods and Figure S1, Supporting Information) that ensures real-time, dexterous, and high-precision operation with a closed-loop accuracy of ~100 nm.[30] Trapping is the basic mode of manipulation, as depicted in **Figure 1**a. When a static 532 nm pulsed laser spot (average power ($P_{ave}$) = 2.5 μW, repetition rate ($f_{rep}$) = 500 Hz, spot diameter ($D_{spot}$) = 5.2 μm) was irradiated on one end of a 7.6 μm-long gold NW (Au NW), collinear but opposite transient thermal gradient forces ($\boldsymbol{F}_{grad}$)[32] were generated inside the NW, driving it toward the laser spot until the NW centroid coincided with the spot center (Figure S2, Supporting Information). By continuously moving the substrate along the axis direction (Figure 1b), the trapped NW underwent repetitive photothermal-shock processes to catch up with the spot and be re-trapped, resulting in axial translation, as shown in Figure 1d. The trapping stiffness ($k$), which quantitatively characterizes the trapping ability of the PTST,[31] was measured to be ~0.60 N m$^{-1}$ within the range of −0.50─0.50 μm where the net force ($F_{Net}$) on the NW is linearly proportional to $x$. Notably, when $F_{Net}$ reaches its maximum value of ~0.36 μN, the NW velocity attains a peak of ~1.9 μm s$^{-1}$ (Figure 1c), implying that the relative moving speed of the substrate and the laser spot cannot exceed this value to avoid the NW detaching from the spot. Therefore, the region corresponding to the maximum velocity −0.85─0.85 μm is the stable trapping region, where the NW will always be trapped in the spot.

For trapped NWs, collinear but opposite $F_{grad}$ and interfacial friction are relatively balanced, but further increasing $P_{ave}$ may break this balance due to the uneven distributions of friction, causing the NW to produce new motions. For thinner NWs, the increased $F_{grad}$ on both sides of the spot is slightly non-collinear, which squeezes the trapped NW away from the spot center to coincide with the iso-intensity line of the spot, as illustrated in Figure 1e (Figure S3, Supporting Information). The lateral translation mode can thus be realized by continuously moving the spot. Additionally, due to uneven substrate surface morphology, a slight angular variation was observed. Figure 1e shows an Au NW was bent and translated laterally ~15 μm by a high-power moving spot ($P_{ave}$ = 3.5 μW, $D_{spot}$ = 5.6 μm). For NWs with small aspect ratios, i.e., for short (length is comparable to $D_{spot}$) and thick NWs, the non-collinear $F_{grad}$ has difficulty bending them and instead pushes them laterally away from the spot center, termed the lateral rotation mode (Figure 1f). For example, a 7.6 μm-long Au NW was rotated ~43° counterclockwise under the influence of a high-intensity moving spot ($P_{ave}$ = 6.5 μW, $D_{spot}$ = 7.1 μm). In addition to Au NWs, the above manipulation was achieved with other NW materials, e.g., silver and palladium NWs (Ag and Pd NWs, Figure S4, Supporting Information).

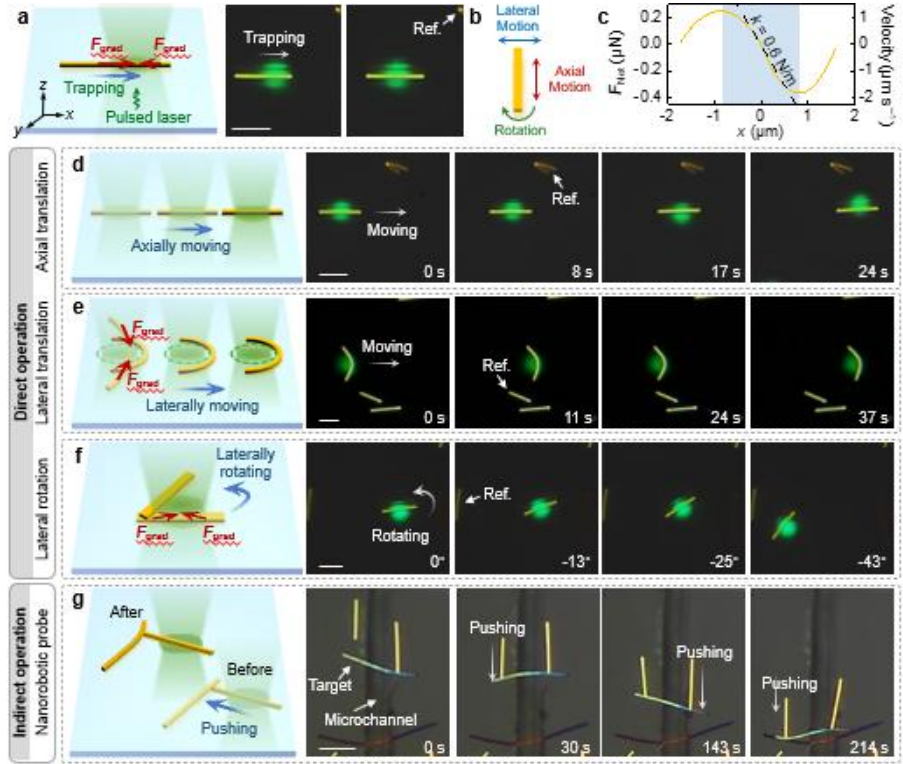

**Figure 1.** Multimodal PTSTs platform. a) A typical trapping process of a 7.6 μm-long Au NW under the static laser spot. b) Diagram of the NW motion directions. c) Deduced $F_{Net}$ and velocity as a function of the *x*-axis coordinate position of the NW from (a). d−f) Direct multimodal operation under the guidance of the dynamic spot, including axial translation, lateral translation, and lateral rotation. g) Typical indirect operation using metal NWs as nanorobotic probes alternately pushing another metal NW across a magnesium fluoride microchannel. Scale bar in all panels: 5 μm.

In addition, the NWs trapped in an axial translation mode can be used as power sources, i.e., nanorobotic probes,[28] to output strong external force to push other micro/nano-objects. Such an indirect operation approach is suitable for manipulating objects with much smaller dimensions or weaker optical absorption, such as quantum dots, microparticles, and non-metallic or special-shaped NWs. As shown in Figure 1g, two Au NWs were guided by a 1064 nm pulsed laser ($P_{ave}$ = 3.5 mW, $f_{rep}$ = 13 kHz, $D_{spot}$ = 5.2 μm) to alternately push a long Ag NW across a magnesium fluoride microchannel, allowing it to be effectively moved from a great distance to the vicinity of pre-placed cadmium selenide (CdSe) NW to construct a hybrid photon-plasmon Mach-Zehnder interferometer (Video S1).

Two-dimensional (2D) materials with excellent electronic and optical properties are anticipated to realize novel optoelectronic devices;[33–35] however, the light-matter interaction significantly constrains their device performances due to their atomic thickness. Integrating 2D materials with plasmonic NWs has been demonstrated to address this issue; however, precise dynamic control over the interaction length between the two remains a technique challenge. Here, we solved this issue easily by using the PTSTs. As shown in **Figure 2**a, a pulsed laser ($P_{ave}$ = 300 μW, $D_{spot}$ = 4.7 μm) with a wavelength of 1064 nm far away from the absorption band of molybdenum sulfide ($MoS_2$) monolayers to prevent material thermal damage, was used as the actuation laser source and translated an Ag NW axially on a $MoS_2$ monolayer (Video S2). Photoluminescence (PL) of the $MoS_2$ monolayer was excited at the same location using a 532 nm continuous-wave (CW) laser ($P_{ave}$ = 4.0 mW, $D_{spot}$ = 4.2 μm) (Figure 2b). As shown in Figure 2c, d, the output intensity of the Ag NW was significantly changed with the interaction length. When the centroid of the Ag NW coincided with the spot center, the output intensity at both ends of the NW was almost equivalent, which may produce attractive potential applications such as valley-controlled directional coupling of light.[36] Furthermore, no additional mechanical scratches were observed on the $MoS_2$ monolayer after NW's translation (Figure 2e, f).

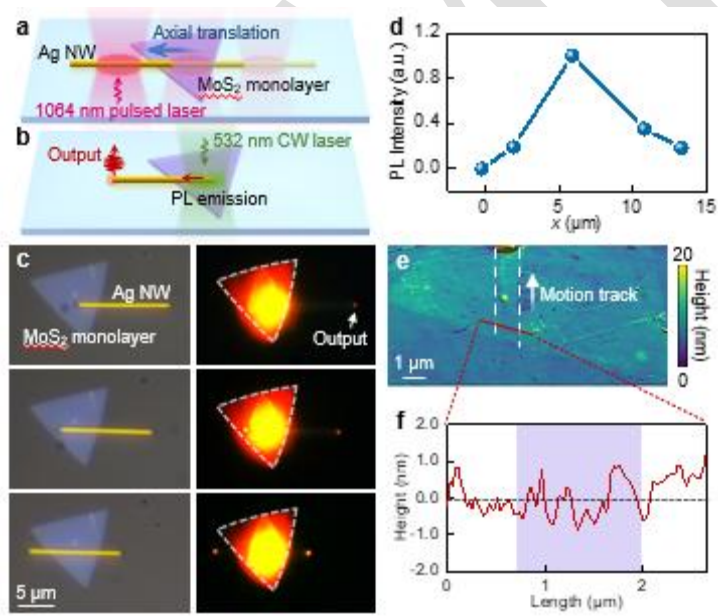

**Figure 2.** Reconfiguration of the interaction length between an Ag NW and a $MoS_2$ monolayer. a,b) Schematic diagrams of the dynamic tuning process of the interaction length and the PL excitation process. A 1064 nm pulsed laser was used as the actuation source in (a), and a 532 nm CW laser for the PL excitation source in (b). c) Optical and PL images of the coupled Ag NW and $MoS_2$ monolayer with varied overlap. d) Output intensity from the Ag NW end as a function of the $x$-axis coordinate position of the Ag NW right end, taking the spot center as the origin. e,f) AFM image of the $MoS_2$ monolayer after NW's translation. Data in (f) are obtained from the random value (along the red line) of the NW motion track in (e).

Semiconductor NWs are widely used as nanolaser sources due to their high optical gain and low lasing threshold. They have been demonstrated to integrate with metal NWs to construct hybrid photonic-plasmonic devices and circuits.[3,4,6] Here, we validated the unique capability of PTSTs to precisely and dynamically adjust the spatial arrangements of discrete components, such as the $d$, $\theta$, and $L$ between the semiconductor and metal NWs. As shown in **Figure 3**a, to avoid damage to the CdSe NW, a 532 nm laser with low $f_{rep}$ of 500 Hz ($P_{ave}$ = 3.5 µW, $D_{spot}$ = 5.2 µm) was employed to initially translate the Au NW axially to one end of a pre-placed CdSe NW, forming an optical end-face coupler. The CdSe NW laser was then excited using the same pulsed laser but with a high $f_{rep}$ of 8 kHz ($P_{ave}$ = 3.5 µW, $D_{spot}$ = 5.2 µm) (Figure 3b). The generated PL emission was coupled into the Au NW through near-field end face coupling and then emitted at its free end. Subsequently, the Au NW was translated axially at a frequency of 30 pulses each time to fine-tune the $d$ between the two NW ends. The $d$ was measured from the AFM images corresponding to the orange box in the optical images of Figure 3c, which exhibited a linear relationship with the pulse number in a range of ~0.6 µm (Figure 3d). The axial motion accuracy was determined to be ~3.9 nm per pulse, substantiating the exceptional control precision and reproducibility of our PTSTs. The PL intensity output at the Au NW free end gradually decreased quasi-exponentially as $d$ increased (Figure 3d, e). When $d$ was beyond ~0.6 µm, the output became imperceptible, mainly due to the low end-face coupling efficiency in the far field. This observed evolution of the end-face coupling from near field to far field on the solid substrate is challenging to technically implement via conventional probe-based techniques due to the lack of axial manipulation capability, which has not yet been reported.

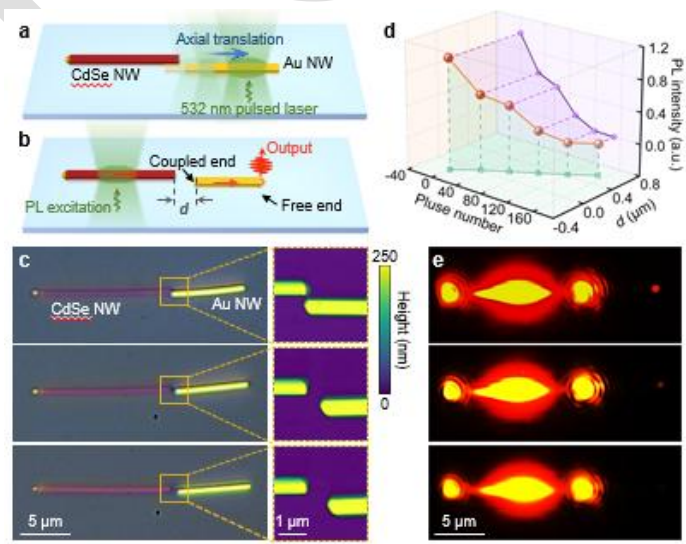

**Figure 3.** Reconfiguration of photon-plasmon coupling distances between a CdSe NW and an Au NW from near field to far field. a,b) Schematic diagrams of the dynamic tuning process of $d$ and the PL excitation process. A 532 nm pulsed laser with a low $f_{rep}$ of 500 Hz was used as the actuation source in (a) and a high $f_{rep}$ of 8 kHz as the PL excitation source in (b). c) Optical and AFM images of the coupled CdSe NW and Au NW as $d$ varied from direct contact to ~0.6 μm, using a frequency of 30 pulses each time to fine-tune the $d$. d) 3D plot of pulse number-dependent $d$ and PL intensity variation with $d$. Note that $d$ exhibits a linear relationship with the pulse numbers while the output PL intensity decays exponentially as $d$ increases. e) Corresponding PL images of the coupled CdSe NW and Au NW with varied $d$.

We further presented dynamic reconfiguration of the $\theta$-dependent photon-plasmon coupling process through a combination of lateral rotation and translation modes. As shown in **Figure 4**a, the low-power 532 nm laser ($P_{ave}$= 4.5 μW, $D_{spot}$ = 5.2 μm) was first employed to actuate an Au NW away from a pre-placed CdSe NW. We then increased $P_{ave}$ to 6.5 μW to rotate the Au NW clockwise. Subsequently, we reduced $P_{ave}$ to the initial 4.5 μW to axially translate the Au NW to its original position near the CdSe NW, forming an angle between the two NWs. Repeating this process increased the $\theta$ from 54° to 91° (Figure 4b). When exciting one end of the CdSe NW, a progressive decrease in output intensity at the Au NW end was observed as $\theta$ increased (Figure 4c, d).

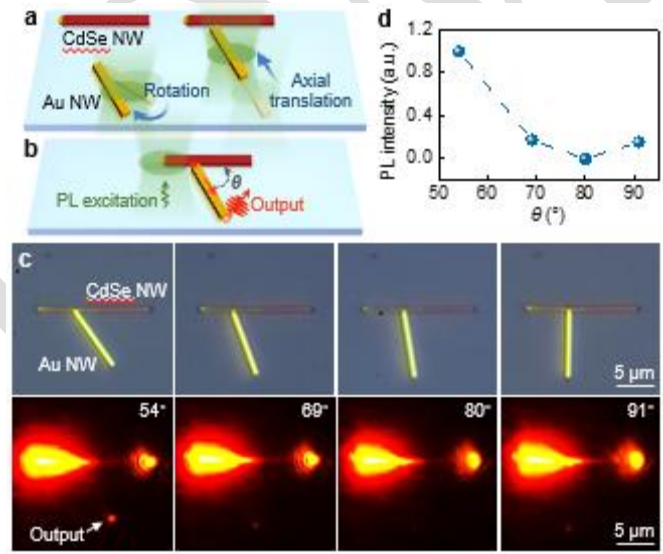

**Figure 4.** Reconfiguration of photon-plasmon coupling angles between a CdSe NW and an Au NW. a,b) Schematic diagrams of the dynamic tuning process of the $\theta$ and the PL excitation process. c) Optical and PL images of the coupled CdSe NW and Au NW with varied $\theta$. d) Output intensity at the free end of the Au NW as a function of $\theta$.

Hybrid photonic-plasmonic cavities have demonstrated outstanding potential in lasers,[3,6] nonlinear optics,[35] surface-enhanced Raman scattering,[37–39] and biosensors,[39,40] where the key lies in achieving strong photon-plasmon coupling within these structures. Therefore, precise spatial adjustment of various cavity parameters, especially the cavity length and *L*, is crucial but difficult to achieve using traditional micro/nanomanipulation. **Figure 5**a displays the schematic diagram of the dynamic fine-tuning of the *L* of a hybrid photon-plasmon cavity. An Au NW was axially translated near the CdSe NW using a 532 nm pulsed laser with low $f_{rep}$ ($P_{ave}$ = 1.5 μW, $D_{spot}$ = 5.2 μm) (Figure 5a I), and then was bent and moved towards the CdSe NW using the lateral translation mode at high $P_{ave}$ of 2.0 μW until the two were closely attached (Figure 5a II), as confirmed by the AFM image at position 3 in Figure S8 (Supporting Information). Subsequently, the Au NW was axially retreated to adjust the *L* (Figure 5a III). The PL emission was excited at the free end of the CdSe NW using the same 532 nm pulsed laser with a high $f_{rep}$ ($P_{ave}$ = 1.5 μW, $D_{spot}$ = 5.2 μm) (Figure 5b).

We observed noticeable changes in the output intensity and spectra at the Au NW free end as *L* was changed (Figure 5c, d). At positions 4 and 3, the short *L* led to weak coupling between the photon and plasmon modes, while the energy coupled into the NW underwent a long SPP propagation distance. In this case, the spectral output at the Au NW free end resembled that at the CdSe NW but with weaker intensity. For CdSe NW itself, its short length of 10.3 μm formed a favorable single-mode Fabry–Perot cavity (Figure 5e), with an average free spectral range (FSR) of 4.79 nm and a dominant peak at 728.92 nm, which has a full width at half-maximum (FWHM) of 0.36 nm and a corresponding quality factor of ~2025.

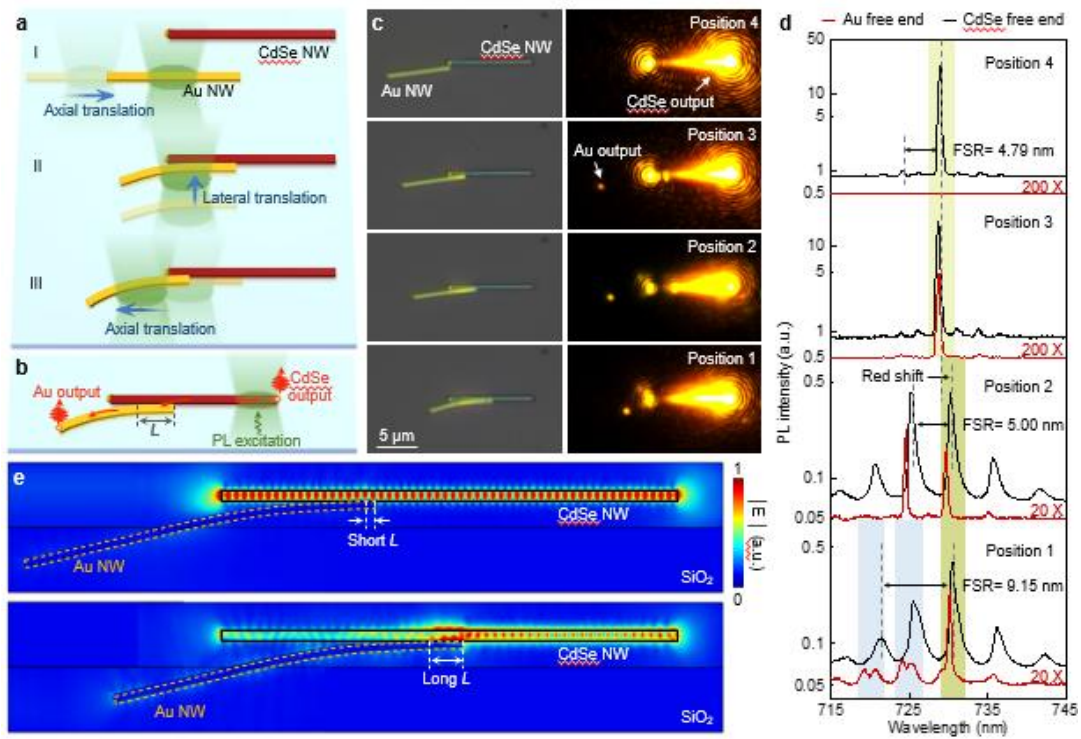

**Figure 5.** Reconfiguration of hybrid photonic cavity-plasmonic cavity coupling. a) Schematic diagrams of the dynamic tuning process of the *L* through 3 typical operations: axial (I), lateral

(II), and axial (III) translation, respectively. b) Schematic diagrams of the PL excitation process. A 532 nm pulsed laser with a low $f_{rep}$ of 500 Hz was used as the actuation source in (a) and a high $f_{rep}$ of 8 kHz as the PL excitation source in (b). c) Optical and PL images of coupled CdSe NW and Au NW with four typical positions. d) Output spectra collected at the Au and CdSe NW free ends. The light green region indicates the dominant peak of the CdSe NW Fabry−Perot cavity around ~729 nm, the dark green region corresponds to the redshifted peak resulting from the increased $L$, and the blue region marks the new split modes. e) Simulated electric field distributions for short and long $L$ conditions.

As $L$ increased (position 2), the Au NW's introduction caused distinct variations in the spectral output of the single-mode Fabry–Perot cavity, including red-shifted peak positions, slightly increased FSR, and significantly broadened FWHM. Notably, the 729-nm dominant peak at the CdSe NW free end gradually weakened and broadened (FWHM = 0.69 nm). The corresponding quality factor decreased from ~2025 to 1048, indicating a severe deterioration of the photonic nature of the resonant cavity. As $L$ further increased (position 1), strong coupling between the photonic and plasmonic modes occurred and formed an efficient hybrid photon-plasmon cavity (Figure 5e, Note S1 and Note S2, Supporting Information), which was reflected by the appearance of new mode splitting at shorter wavelengths around 719 nm and 724 nm (in blue region) and a significant increase in FSR from ~4.79 to 9.15 nm. It is worth mentioning that both the quality factor measured at the free ends of CdSe and Au NWs increase to ~1404 and 1765, respectively, signifying the high quality of the hybrid photonic-plasmonic cavity.

Finally, we demonstrated the indirect nanorobotic probe-based approach to fine-tune semiconductor NWs and special-shaped metal NWs to achieve optimal performances of plasmonic nanostructures. As shown in **Figure 6**a, a 1064 nm pulsed laser ($P_{ave}$ = 3.0 mW, $f_{rep}$ = 13 kHz, $D_{spot}$ = 5.2 μm) was employed to drive an Au NW as a nanorobotic probe to adjust the posture of the CdSe NW in a pre-fabricated hybrid photon-plasmon cavity, which is barely absorbed and cannot be manipulated directly with 1064 nm light. The PL emission was excited at the middle of the CdSe NW using the 532 nm pulsed laser ($P_{ave}$ = 3.0 μW, $D_{spot}$ = 5.2 μm) (Figure 6b). Thanks to nanoscale stepping precision and dexterity, the resonant modes were easily changed and consequently realized the conversion of the output lasing from multi-mode to single-mode, with a side-mode suppression ratio considerably improved from 0.9 dB to 15.7 dB. The boost in the intensity of the dominant single mode considerably narrowed the FWHM from ~0.41 to 0.25 (Figure 6c). Figure 6d and e display the reconfiguration of a 3-dB plasmonic coupler by using an Au NW nanorobotic probe to push a special-shaped Au NW to couple with another Au NW in a pre-fabricated hybrid photon-plasmon cavity. Through fine-tuning, the output ratio at both output ends was ultimately converted from 5:1 to 1:1.

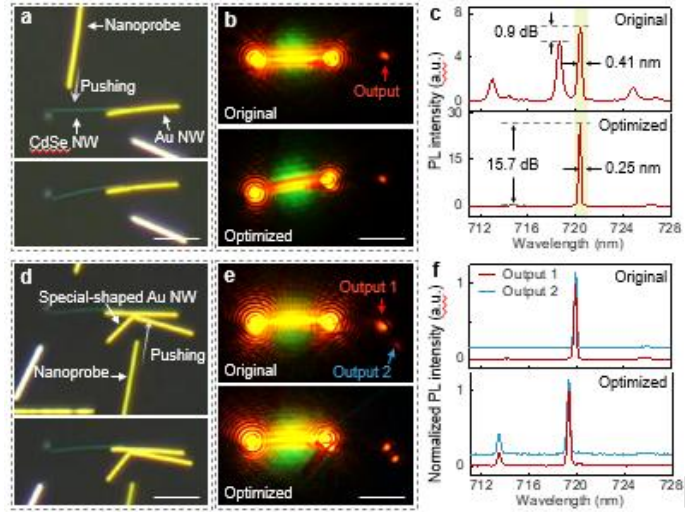

**Figure 6.** Structure reconfiguration and performance optimization by using the nanorobotic probe-based indirect operation mode. a) Single-mode hybrid photon-plasmon laser reconfiguration process. An Au NW nanorobotic probe is used to push the CdSe NW in a pre-fabricated hybrid photon-plasmon cavity. b) Schematic diagrams of the PL excitation process, and c) the corresponding spectral output collected at the Au NW free end. d) 3-dB plasmonic coupler reconfiguration process. An NW nanorobotic probe is used to push the long side of a special-shaped Au NW close to the pre-placed Au NW in a pre-fabricated hybrid photon-plasmon cavity. e) Schematic diagrams of the PL excitation process, and e) the corresponding spectral output collected at the two Au NW free ends. Scale bar in all panels: 5 μm.

## 3. Conclusion

In summary, we have successfully demonstrated the in-situ dynamic spatial reconfiguration of plasmonic NW devices and circuits on dry solid substrates, which thoroughly showcases the remarkable advantages of the PTSTs platform in manipulating single NWs, including versatility, dexterity, nanoscale high precision, real-time operation, and large external output force. This allows us to easily observe some experimental phenomena that have not been reported before or are challenging to achieve with traditional micro/nanomanipulation, such as the intensity evolution of axial photon-plasmon coupling from near field to far field, and the resonant mode evolution of photonic cavity-plasmonic cavity coupling from weak to strong. For metallic nanowires, we can use the direct operation mode, while for metallic and dielectric nanomaterials with smaller sizes, weaker light absorption, or unique shapes, we can use the nanorobotic probe-based indirect operation mode. Combining the two makes it possible to achieve spatial position manipulation of almost any size, structure, and material type. In addition to 2D planar substrates, our results also apply to other 1D substrates, such as the narrow planes of lithographically fabricated micro/nano-waveguides[41–43] and the curved surfaces of thermally stretched micro/nanofibers.[10,32] In the future, by combining artificial intelligence (AI) technologies[44,45] such as machine vision and deep reinforcement learning, an intelligent robotic PTST system can be built to achieve fully autonomous, high-throughput in-situ spatial reconfiguration, spectral detection, and intensive data analysis of plasmonic devices and circuits, which is expected to promote the establishment of a new research paradigm in the field of nanoplasmonics.[46]

## 4. Methods

*Materials Fabrication and Preparation:* Single-crystal metal NWs were synthesized by a thermal evaporation method,[28] specifically, Au NWs were grown at 1250°C for 40 to 120 minutes, Ag NWs at 1050°C for 60 to 90 minutes, and Pd NWs at 1350°C for 2 hours. CdSe NWs were synthesized using gold as a catalyst in a standard chemical vapor deposition process at 830°C for 30 to 60 minutes. Additionally, $MoS_2$ monolayers were synthesized on a silicon substrate using the chemical vapor deposition method and then transferred to a fused silica substrate via a wet transfer process.[47,48]

*Multimodal PTSTs Platform:* The platform consists of three parts: a PTSTs system, an integrated control module, and an image feedback module.[30] The PTST module is the key part for trapping and manipulating samples. A 1064 nm nanosecond pulsed laser and its frequency-doubled 532 nm laser were used as the actuation light sources. Unless otherwise specified, the $f_{\text{rep}}$ of the actuation source in the experiment was 500 Hz. The actuation lasers were focused on the target from the bottom through a 50× objective (NA = 0.55). $D_{\text{spot}}$ could be changed at the focus plane by tuning a piezo *z*-stage. A galvanometer system (GVS002, Thorlabs) and a piezo *xy*-stage (NPXYZ100SGV6, Newport) were employed to adjust the sample location for manipulation. A camera recorded the samples through a 100× objective (NA = 0.8). The integrated control module, the core part for controlling the PTSTs, consists of an embedded processor (ZYNQ-7000, Xilinx), a driving circuit, and a computer for integrated intelligent control. The image feedback module established the closed-loop feedback based on image vision. Figure S1 in Supporting Information provides more details of the experimental setup.

*Morphology Characterization:* Material morphology profiles in Figure 2, 3, and S8 were measured by a high-resolution AFM (Cypher S, Oxford Instruments Asylum Research) operating in the tapping mode (with a uniform scan rate of 0.5 Hz).

*Spectral detection:* The PL emissions of the MoS$_2$ monolayer were excited by a 532 nm CW laser from top to bottom, while the lasing of CdSe NWs or hybrid cavity was excited by a 532 nm nanosecond laser from bottom to top. Unless otherwise specified, the $f_{rep}$ of the 532 nm pulsed excitation source was 8 kHz. The PL intensity in Figure 2, 3, and 4 was obtained by first collecting PL images using a CCD, and then processing them with commercial software to integrate the intensity within the specified position and range in the pictures. The spectra presented in Figure 5 and 6 were acquired by a spectrometer (iHR550, HORIBA, Ltd.) under a 50× (NA = 0.55) or 100× (NA = 0.8) objective, respectively.

*Motion Detection and Image Analysis:* The target NW movement's translation distances and rotation angles were determined using an AI image recognition method (YOLOv8 algorithm). We developed an automated image recognition program to extract the locomotion information for target NW objects. The boundaries of each target are delineated by extracting the grayscale values from real-time images, which can enhance the recognition capability and allow for precise tracking of NWs. The tracking accuracy depends on the resolution of the captured images, which corresponds to ~84.1 nm pixel$^{-1}$ for the recorded videos in our experiments. Furthermore, calculating the average single pulse step is based on measuring the translation distance of the NW centroid over several hundreds or thousands of laser pulses. Determining the averaged single-step size is simplified to measure the accumulated centroid locomotion distance divided by the input pulse number. Therefore, the single-step accuracy of this method will improve as the pulse number on the target sample increases.

## Supporting Information

Supporting Information is available from the author.

## Acknowledgements

This work is supported by the National Natural Science Foundation of China (grant nos. 62122054, 62075131, and 12304333). and Guizhou Provincial Basic Research Program (Grant No. ZK[2024]071).

## Author Contributions

F.G. conceived the idea and supervised this project. R.Z. performed experiments and data analysis. Z.G. and T.S. performed the simulations. F.G. and R.Z. co-wrote the paper. S.Z., Y.L., and S.Linghu. provided many insightful suggestions. All authors discussed the results and commented on the manuscript.

**Conflict of Interest**

The authors declare no conflict of interest.

**Data Availability Statement**

The data that support the findings of this study are available from the corresponding author upon reasonable request.